\documentclass[epj]{svjour}
\usepackage{graphics}
\usepackage{hyperref}
\hypersetup{colorlinks = true, allcolors = blue}
\usepackage[utf8]{inputenc}
\usepackage{authblk}
\usepackage{hepparticles}
\usepackage{hepunits}
\usepackage{hepnames}

\begin{document}
\title{Exploring requirements and detector solutions for FCC-ee}
%\subtitle{If you have a sub-title, enter it here}
\author{Patrizia Azzi\inst{1} \and Emmanuel Perez\inst{2} % etc
}                     % Do not remove
\offprints{}          % Insert a name or remove this line
\institute{INFN, Sezione di Padova, Italy \and CERN, EP Department, Geneva, Switzerland}
\date{
{\sl 
(Submitted to EPJ+ special issue:
A future Higgs and Electroweak factory (FCC): Challenges towards discovery, Focus on FCC-ee)}
%Received: \today / Revised version: \today 
}
\abstract{
%Insert your abstract here.
Circular colliders have the advantage of delivering collisions to multiple interaction points, which allow different detector designs to be studied and optimized – up to four for FCC-ee. On the one hand, the detectors must satisfy the constraints imposed by the invasive interaction region layout. On the other hand, the performance of heavy-flavour tagging, of particle identification, of tracking and particle-flow reconstruction, and of lepton, jet, missing energy and angular resolution, need to match the physics programme and the exquisite statistical precision offered by FCC-ee. During the FCC feasibility study (2021-2025), benchmark physics processes will be used to determine, via appropriate simulations, the requirements on the detector performance or design that must be satisfied to ensure that the systematic uncertainties of the measurements are commensurate with their statistical precision. The usage of the data themselves, in order to reach the challenging goals on the stability and on the alignment of the detector, in particular for the programme at and around the Z peak, will also be studied. In addition, the potential for discovering very weakly coupled new particles, in decays of Z or Higgs bosons, could motivate dedicated detector designs that would increase the efficiency for reconstructing the unusual signatures of such processes. These studies are crucial input to the further optimization of the two concepts described in the Conceptual Design Report, CLD and IDEA, and to the development of new concepts which might actually prove to be better adapted to the FCC-ee physics programme, or parts thereof.
%
%\PACS{
%      {PACS-key}{describing text of that key}   \and
%      {PACS-key}{describing text of that key}
%     } % end of PACS codes
} %end of abstract
\maketitle

\section{Introduction}
\label{section:intro}
Detectors for FCC-ee  have to comply with the tight constraints imposed by the invasive machine detector interface~\cite{Abada2019}: the last focusing quadrupole is at two meters only from the interaction point (IP); the experiment magnetic field is constrained to be below $2$\,T for the run at the Z resonance; the angular coverage of the detector can not extend below $100$\,mrad from the beam axis, as this space is used by machine magnets. The experimental environment, which is different from that of a linear collider in particular at the Z peak (high physics event rates, small bunch spacing), also sets important constraints, preventing for example the use of pulsed electronics. The detector requirements imposed by the physics programme~\cite{Abada:2019lih,Blondel:2021ema}, at 240\,GeV and above, have already been studied extensively for the linear colliders, but will have to be revisited in the context of the FCC-ee environment. Moreover, the huge statistics anticipated at the Z resonance (the so-called ``Tera-Z" run) comes with specific  challenges, as the systematic uncertainties of the measurements should be commensurate with their very small statistical uncertainties. In addition, the specific discovery potential for very weakly coupled particles, offered by the huge FCC-ee statistics, should be kept in mind too when designing the detectors.

\section{Control of acceptances}

One of the strongest requirements that is imposed by the Tera-Z programme concerns the determination of the acceptances, which, generally, have to be known with an accuracy in the range from a few $10^{-6}$ to $10^{-4}$. For example, for the luminosity measurement~\cite{R1lumi}, the goal is to reach an uncertainty of $10^{-4}$ from low-angle Bhabha events, which would match the anticipated theoretical precision on the Bhabha cross section. With the luminosity monitor (LumiCal) at $1$\,m only from the interaction point, and the measurement starting at an angle of $65$\,mrad, the inner radius of the LumiCal must be known to $1.6\,\mu$m only~\cite{Abada2019}. 
A second example is provided by the measurement of $R_\ell$, the ratio of the hadronic to leptonic Z decays, for which the lepton acceptance is a key systematic uncertainty. To match the anticipated statistical uncertainty, a careful control of the boundaries is again mandatory. For example, 
%for a measurement at $\theta$ above 15 degrees, this angular boundary must be known to less than 8 $\mu$rad, which means that the innermost radius of the endcap calorimeter should be known to 15 $\mu$m.
the innermost radius of the endcap calorimeter should be known to ${\cal{O}}(15)$\,$\mu$m, which poses constraints on the mechanical assembly of the calorimeter. A hermetic calorimeter would be better suited than a petal design in this perspective.

\section{Measurement of the tracks of charged particles}

\paragraph{Angular resolutions}
A precise determination of the beam energy spread (BES) is crucial for several key measurements at FCC. In particular, the BES affects the Z lineshape and it would have a huge effect on the extracted Z width if unattended. As accelerator diagnostics cannot provide the required level of precision, the BES has to be determined by the experiments, and the method traditionally used at LEP, exploiting the size of the luminous region, can not be applied at FCC because of the crossing angle. However, it can be measured at the level of a few per-mil from the scattering angles of dimuon events~\cite{blondel2019polarization}. Because of the constrained kinematics of such events, the longitudinal imbalance can be reconstructed event by event, and the BES can be extracted from the width of its distribution. To ensure that the BES has a negligible effect on the extracted Z width, muon tracks from Z decays must be measured with an  angular resolution of 0.1\,mrad or better, a requirement that is fulfilled~\cite{Bacchetta:2019fmz} by the detector concepts presented in the CDR.

\begin{center}
\begin{figure*}
%\vspace*{5cm}       % Give the correct figure height in cm
\center{\resizebox{0.75\textwidth}{!}{
 \includegraphics{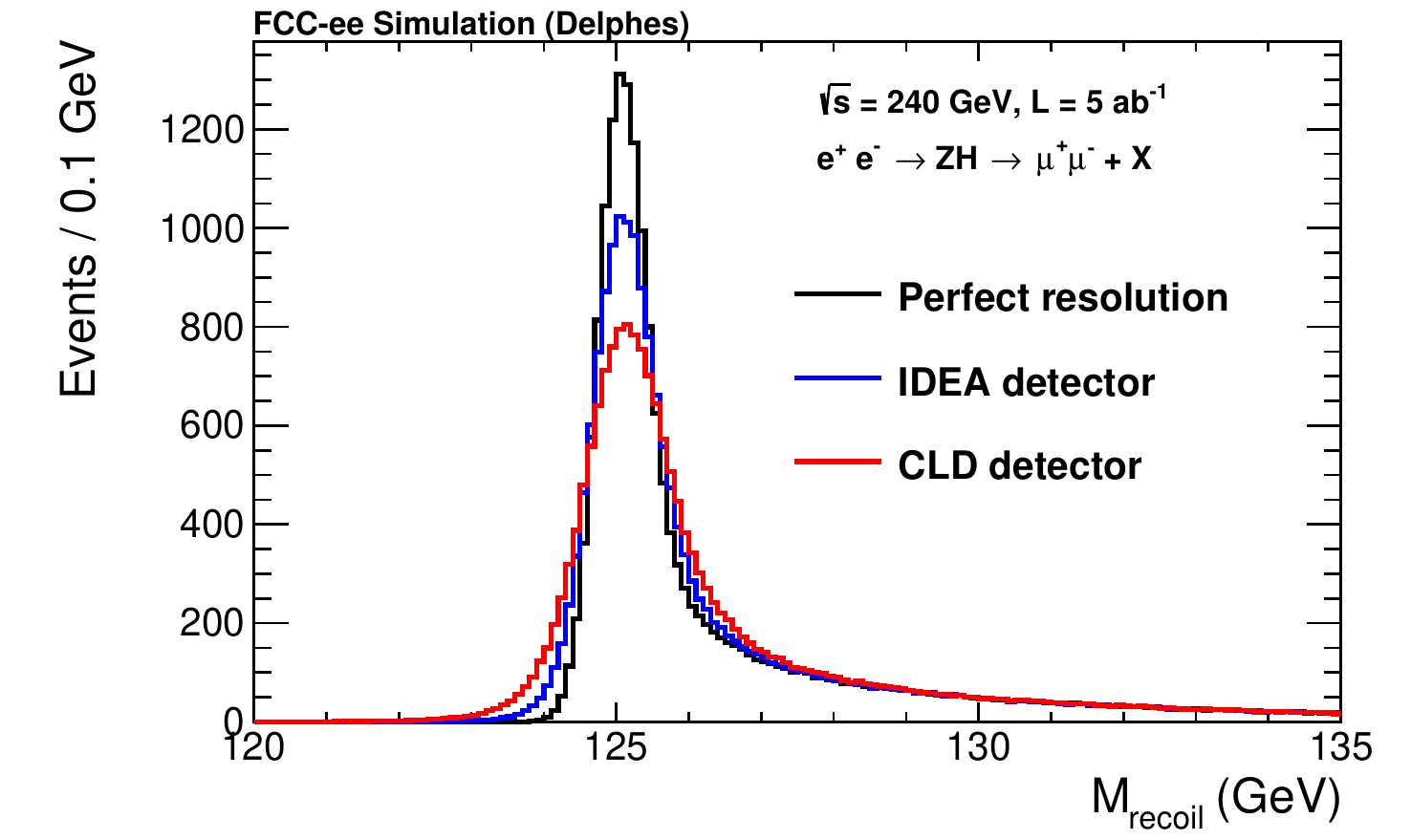}
}
}
\caption{Distribution of the Higgs recoil mass in ZH events where the Z decays into muons, assuming: (black) an ideal momentum resolution, such that the resolution on the recoil mass is determined by the beam energy spread; (blue) the momentum resolution of the IDEA detector; (red) that of the CLD detector. }
\label{fig:mH-recoil}       % Give a unique label
\end{figure*}
\end{center}

\paragraph{Momentum resolution}
The beam energy spread, at the Z resonance and at the ZH energy, also sets a target for the track momentum resolution. This is illustrated in Fig.~\ref{fig:mH-recoil}, which shows the Higgs recoil mass in ZH events where the Z decays into muons. The goal is that the reconstruction of the recoil mass be limited by the BES and not by the detector resolution. The very light tracker from IDEA, with a resolution of ${\cal{O}}(0.15 \%)$ for central, $50$\,GeV muons, is close to reaching this goal. The (heavier) full silicon tracker of CLD performs a bit worse because, in the momentum range of interest, the resolution is dominated by multiple scattering. The determination of the Higgs mass (for which a precision of a few MeV would be needed in view of a possible run at the Higgs resonance) will clearly benefit from the better momentum resolution offered by a light, gaseous tracker~\cite{Azzurri:2021nmy}. 
%Similar requirements that the momentum resolution be comparable to the BES come from physics at the Z peak too. 
A momentum resolution comparable to the BES for beam-energy muons is also important for Z physics. 
For example, the analysis strategy for searching for Lepton Flavour Violating Z decays into $\tau \mu$ demands a clear tau decay in one hemisphere, and a beam-energy muon in the other, in order to suppress the ${\rm Z} \rightarrow  \tau \tau$ background: the sensitivity improves linearly with the momentum resolution on the muon~\cite{Dam:2018rfz}. A precise measurement of the $\tau$ mass will also put some constraints on the track momentum resolution. Requirements are also expected from flavour physics, where momentum resolution is often a key for reducing the backgrounds. On the other hand, the measurement of the Higgs coupling to muons is unlikely to be a good benchmark process for determining this requirement: because of the very low statistics, a resolution $4$ times better than the exquisite one assumed in Ref.~\cite{An:2018dwb}  would be needed, in order to barely reach the precision anticipated at HL-LHC on this coupling.

\paragraph{Stability of the track momentum scale} As shown in Ref.~\cite{blondel2019polarization}, a control of the point-to-point uncertainty on the centre-of-mass energy in the lineshape scan, at the level of $40$~keV, can be achieved in situ from the invariant mass distribution of dimuon events. Such a precision demands that the scale of the momentum measurements, and in particular the detector magnetic field, be stable at (or that its variations be monitored to) the level of $40$~keV / $91$~GeV, hence to a few $10^{-7}$. Such a precise monitoring may be difficult to achieve with magnetic NMR probes, but the large statistics of usual resonances ($J/\psi \rightarrow \mu \mu$, ${\rm D}^0 \rightarrow {\rm K} \pi$, etc) may provide an in-situ monitoring down to this challenging precision.

\section{Requirements on the vertex detector}
The requirements on the resolutions on the track impact parameter are currently believed to be similar to, or better than, those derived from Higgs studies performed in the context of the ILC or CLIC, typically $\sigma = a \oplus b / \sin^{3/2} \theta$ with $a \simeq 5\,\mu$m and $b \simeq 15\,\mu$m. These requirements will have to be reached despite the additional constraints set by the FCC environment on the readout electronics of the detector: (i) the power budget is smaller than for a detector operating at a linear collider (since power-pulsing the electronics is not possible with collisions occurring every $\sim 20$\,ns); and (ii) it should be fast enough, less than about 1 $\mu$s, such that the integrated background remains negligible. In addition to the measurement of the Higgs couplings to pairs of b quarks, c quarks and gluons, which demands a high-performance flavour tagging, requirements on the vertex detector will come from the measurement of heavy-quark electroweak observables ($R_{\rm b}$, $R_{\rm c}$ and the heavy-quark forward-backward  asymmetries), for which a huge improvement is expected compared to LEP, benefiting not only from the large luminosity increase, but also from decades of improvements in detector technology, which, currently, lead to b-tagging efficiencies that are three times larger than those achieved at LEP for the same mis-tag efficiency. Moreover, the rich flavour physics programme at the Z pole is expected to set demanding goals on the resolutions with which vertices are reconstructed. 
For example, an excellent vertexing is a key for extracting a signal of $\rm B \rightarrow K^* \tau \tau$ when both $\tau$s decay into three charged pions, as it allows this decay to be fully reconstructed. Assuming resolutions of $3\,\mu$m, $7\,\mu$m and $5\,\mu$m respectively, for the reconstruction of the primary vertex, of the B decay vertex, and of the $\tau$ decay vertices, more than one thousand events will be reconstructed~\cite{Abada2019}, opening the door for measurements of the angular properties of this decay, which is likely to be unique to FCC.

%\section{Energy resolution of the electromagnetic calorimeter}
\section{Requirements on the electromagnetic calorimeter}

Since about $25 \%$ of the jet energies is carried by photons, a good energy resolution for photons is needed for a good measurement of jets. The corresponding requirement is however not very stringent: a stochastic term of 15--20\% is  enough to ensure a jet energy resolution better than $3\%$ for $50$\,GeV jets, using a particle-flow reconstruction algorithm~\cite{Lucchini:2020bac}. 
The measurement of the Higgs to $\gamma \gamma$ coupling obviously benefits from a good electromagnetic resolution. This measurement is however very statistically limited at FCC-ee. With a resolution of $\sigma(E)/E = 15\% / \sqrt{E} \oplus 1 \%$, the anticipated precision on the  $\rm {H} \gamma \gamma$ coupling amounts to ${\cal{O}} ( 3 \%)$~\cite{An:2018dwb}, a factor of two worse than what HL-LHC should achieve. Even with an exquisite stochastic  term and a constant term well below $1 \%$, it will be difficult to achieve a precision significantly  better than  that of the anticipated HL-LHC measurement. On the other hand, requirements of a resolution much better than $15 \% / \sqrt{E}$, in particular at rather low energies, are expected from flavour physics, as many important measurements of CP violation rely on the reconstruction of decays with several $\pi^0$'s in the final state, as in $B^0 \rightarrow \pi^+ \pi^- \pi^0 \pi^0$. The extraction of a $B_s \rightarrow D_s K$ signal in modes with neutral pions is likely to  require an exquisite resolution of $ 5\% / \sqrt{E}$ or better, as  can be achieved with crystal calorimetry~\cite{aleksan2021cp}. Such a resolution is also required for a  precise measurement of  the ${\rm Z} \nu_e {\bar{\nu}}_e$ coupling, exploiting radiative return events with a single photon in the final state~\cite{Aleksan_2019}. Moreover, the electromagnetic resolution is a key for pushing the sensitivity to rare  or forbidden processes, like  the  $\tau  \rightarrow \mu \gamma$ or ${\rm Z}  \rightarrow \tau e$ decays~\cite{Dam:2018rfz}.  
%Also: EM resolution could be important for BSM signatures from events with ALPS with multiple photons in the final state. 
%%%%
%An exquisite resolution would also allow a performant recovery of photons radiated by electrons – with a stochastic term of $5\%$, the resolution of the Higgs recoil mass in the  channel ${\rm Z} \rightarrow  ee$ would get close to that obtained with muons, such that the  electron channel would contribute significantly to the precise determination  of the Higgs mass. 
%The jet resolution may also be improved by clustering photons into $\pi^0$’s prior to clustering the jet~\cite{Lucchini:2020bac}.
In addition, the jet resolution may also benefit from an exquisite electromagnetic resolution, that would allow photons to be clustered into $\pi^0$’s prior to clustering the jet~\cite{Lucchini:2020bac}. \\

A high granularity of the electromagnetic calorimeter   plays a crucial role in the identification of individual photons in jets, coming for example from the decay of $\pi^0$'s or low mass axions or axion-like particles (ALPs), and, more generally, is a key for an optimal particle-flow reconstruction. Requirements on the granularity will be studied using as benchmarks the measurement of the tau polarisation, and the sensitivity on low mass ALPs, that could be copiously produced in Z decays~\cite{Bauer:2018uxu,steinberg2021axionlike}.

\section{Jets and resolution of hadronic systems}

Among the many advantages of an electron-positron collider such as the FCC-ee, the most cited ones are the cleanliness of the final state and the precise knowledge of the centre-of-mass energy. This comes in stark contrast with the situation at an hadronic collider, such as the LHC for example, where for many measurements and searches the analysis needs to recur to the selection of leptonic final states to control backgrounds and uncertainties, sacrificing a large fraction of the event statistics. 
At a lepton collider instead, the hadronic final states are  very important players in the overall physics program, often driving the measurement statistical precision for many (sometimes rare)  processes. 
Uncertainties on the jet properties directly propagate to the measurement precision and need to be understood and controlled in order to achieve a high precision or a better background rejection.

A jet is a complex object that derives from the clustering of the products of the fragmentation process of an hadronic particle. Optimal reconstruction approaches, such as particle-flow, use information from all sub-detectors, beyond just the one from the calorimeter, imposing requirements on the design of the overall detector. 

A very important aspect, such as the jet energy resolution is affected both by the choice of the algorithm but also by the intrinsic fluctuation of the fragmentation process itself, and cannot be fully disentangled. For this reason, it is better to exploit the resolution not of a jet, but of a color singlet object such as a W, Z, or Higgs boson, and use the mass of the particle itself as a well defined quantity to assess detector performance. In the case of a lepton collider, this variable could also be not a specific particle but the visible mass of the event or the missing mass. 
In addition, the jet energy resolution might be less critical in certain situations when kinematical constraints can be fully exploited from the precisely known beam energy and particle masses, for instance when no missing energy is present. It is the angular resolution, less affected by the effects quoted above, that once used in the kinematical fits can provide a more robust estimate of the energy, even more so if the system is boosted. 
Trying to factorise the requirements imposed by the need of precise jet reconstruction on the detector, it has to be noted that in the case of a particle-flow reconstruction, the most important characteristic for the calorimetry is not just its energy resolution but the granularity that allows the best matching with the tracker information and the best identification capabilities for neutral particles (photons and neutral hadrons).  
%[Would like to add a plot here maybe]
Once the list of particle candidates is defined, they need to be clustered into jets objects. 
The current set of jet clustering algorithms available is still very much similar to the one used in the past at LEP. They might need to be re-optimised considering that with the new technologies for future experiments the limit would not be the detector resolution, but the focus should be instead on the perfect assignment of the particle candidates to the jets, and then of the jets to their parent particle.  
The new algorithms could also be informed by progress on the theoretical side on QCD and jets substructure to improve the accuracy of the jet definition. 
The ambiguity of the assignment could end up being one of the dominant systematic in multi-jet final states (four of more jets such as ZH or $\rm t\bar t$ events). 
Thinking ahead of the possibilities for future algorithmic development we should take the opportunity of developing a global event reconstruction having all the particle candidates fed into a Deep Neural Network that would take care of associating them, without having an intermediate clustering step. An optimization of this approach could imply a different strategy for the optimization for the detector design as well. 
%[Cite MODE Collaboration once the preprint is published??] 

In practical terms, the quality of the jet energy resolution can be relevant in those events where the kinematics are not over constrained, such as in the case of hadronic decays accompanied by missing energy, or when there is a need for a strong background rejection in the early stage of an analysis. 
For instance, the ILC/CLIC studies derived that a resolution of $\sigma (E_{\rm jet}) / E_{\rm jet} \simeq 30\% / \sqrt{E_{\rm jet}}$ is needed to separate Z and W in their hadronic decays, which is valid irrespective of the overall event environment. This statement can possibly be challenged and the requirements reviewed, in the case of the FCC-ee, for those cases where fits can be used to constrain the kinematics of fully hadronic final states in specific processes, such as ZH production. 
However, it is clear that examples exist where the hadronic resolution on the missing mass can be a very important tool to distinguish between similar processes, such as the separation of  $\rm e^+e^-\to H\nu\nu$ (via WW fusion) from  $\rm e^+e^- \to ZH$ with  $\rm Z\to \nu \bar{\nu}$, instrumental for the determination of the Higgs width.

\section{Particle identification}
Excellent lepton and photon identification capabilities are essential for many analyses. In particular, a good separation $\rm e / \gamma$, $\gamma / \pi^0$, $\rm e / \pi$, and an excellent separation of photons from neutral hadrons, are key ingredients for an effective particle-flow reconstruction. This separation should remain effective also in collimated topologies, as required by a precise measurement of the $\tau$ polarisation for example. 
Moreover, charged particle identification will be mandatory to the flavour physics programme. A separation of pions from kaons, in a momentum range that extends up to at least $10$\,GeV, will be vital for time-dependent CP violation measurements. Separation at higher momentum would be extremely useful too; for example, the spectrum of the kaon in the $\rm B_s \rightarrow D_s K$ decay, a process that suffers from an order-of-magnitude larger background from $\rm B_s \rightarrow D_s \pi$, extends up to $\sim 30$\,GeV. The precise determination of the branching ratios of the tau, and of the tau polarisation, will also benefit from a separation of pions from kaons  up to $\sim 40$\,GeV.
%to suppress background contamination in many flavour physics analyses. First studies have been made in the context of the  $B_s \rightarrow D_s K$ decay, which suffers from a large background from e.g. $B_s \rightarrow D_s \pi$. A separation of pions from kaons, in a momentum range that extends up to at least $10$~GeV, will be mandatory, with a level of separation that will be determined by detailed studies of this and similar modes. Separation at higher momentum, ideally up to $30-40$~GeV, would be very useful too, for example for tau physics. 
Candidate technologies are being reviewed. With the IDEA drift chamber, the ``cluster counting" method looks promising and may cover the whole momentum range of interest. For a detector with a full silicon tracker, no ideal solution exists yet, as it is not easy to cover the whole momentum range and, at the same time, comply with the space and hermiticity constraints of the experiment.

\section{Detection of new, long-lived particles} 

Some new physics processes can produce very long-lived particles (LLPs) that decay far from the primary interaction point, producing a secondary interaction vertex, containing charged and neutral SM particles.  
Other exotic models might produce particles that would give rise to “short", “broken”, or vstopped" tracks signatures. In addition, we could expect also more complex unusual signatures such as “emerging" or “dark showers" jets. A review of the possible models that have been considered can be found in Ref.~\cite{chrzaszcz2021hunt} (see also~\cite{Abada:2019lih,Alipour-Fard:2018lsf,Antusch:2016vyf,Bauer:2018uxu,Wang:2019orr,Blondel:2014bra}).
%Section-"Hunt for rare processes and long-lived particles at FCC-ee" (see also  (see also~\cite{Abada:2019lih},\cite{Alipour-Fard:2018lsf} \cite{Antusch:2016vyf},\cite{Alipour-Fard:2018lsf} \cite{Bauer:2018uxu},\cite{Wang:2019orr}  ,\cite{Blondel:2014bra}).

These peculiar experimental signatures are very distinct from those of Standard Model processes and, if observed, they would be a clear sign of new physics. 
Unfortunately, at colliders standard trigger and reconstruction techniques are typically unable to recognise them efficiently. 
The large statistics and the clean environment of the FCC-ee Tera-Z run  makes it the ideal playground to optimise these types of searches and it has been shown that it can be competitive and complementary, in mass and coupling range, to the discovery reach of non-collider experiments~\cite{Beacham_2019}.  
The needs for an efficient detection of such signatures might call for the proposal of a specific detector with optimised design choices in addition to the improvement of reconstruction techniques of more general purpose ones. 
The variety of signatures imposes requirements on several different experimental aspects. 
First and foremost a very large, light and granular tracking system, that would allow to reconstruct efficiently charged tracks, characterised by potentially having a short/variable length and starting significantly away from the primary collision vertex. 
This could be extended to a choice of technology for the calorimeter that would allow to disentangle also emerging or dark shower jets, that might start outside the tracker radius. 

Timing information will be also essential to be sensitive to heavy particles, with $\beta < 1$,
leading to out-of time or even stopped decays. 
Once appropriate design and technology choices guarantee the detection of these particles signatures, most of the effort has to be spent in the optimisation and development of new reconstruction algorithms. 
The requirement on track and vertex reconstruction, with the ability to identify and measure the charge and momentum of tracks with a small number of hits, and large displacements, and the possibility of reconstructing (multiple) vertexes significantly displaced in the tracking volume and containing a small number of tracks. The requirement on jet reconstruction to be able to identify multiple sub-components, with fewer particles that normal hadronic jets and possibly starting in the calorimeter at different depths. 
This is a case where new Machine Learning techniques could be employed, profiting of the particle-flow event reconstruction that allows to feed the networks with more granular information, such as the particle candidates themselves. 
Finally, another challenging aspect is the similar effort to identify and reduce or control the backgrounds to these unusual signatures. These backgrounds are generally dominated by machine, detector or other external factors, and tend to require specific techniques to be evaluated and monitored. In this respect, there is a role to play for the data acquisition system to profit from both in-time and off-time information  for the study of potential asynchronous signal and backgrounds.

\section{Conclusions}
\label{section:conclusion}
An initial list of the requirements that a FCC-ee detector should fulfil, in order to match the
physics programme offered by the huge statistics that will be collected, has been established. A first list of benchmark processes, that will allow additional requirements to be quantified and better defined, is identified. They will be studied carefully in order to complete the ``wish-list" of detector requirements. The implementation of these requirements into detector designs is likely to come with compromises - for example, a dedicated detector for particle identification, in front of the calorimeter, would unavoidably degrade the electromagnetic resolution. The possibility of having four interaction points offers very interesting possibilities, as complementary detectors could be designed in view of ideally covering the whole FCC-ee physics programme.

%
% For  figures use
%\begin{figure*}
%\vspace*{5cm}       % Give the correct figure height in cm
%\includegraphics{leer.eps}
%\end{figure*}
% or  this
%\begin{figure}
%\centering
%\resizebox{0.75\textwidth}{!}{%
%  \includegraphics{leer.eps}
%\vspace{5cm}       % Give the correct figure height in cm
%\end{figure}
%\begin{table}
%\centering
%\begin{tabular}{lll}
%\hline\noalign{\smallskip}
%\end{tabular}
% Or use
%\vspace*{5cm}  % with the correct table height
%\end{table}

\bibliographystyle{myutphys}
\bibliography{references}
\end{document}